\documentclass[prd,amsmath,amssymb,floatfix,superscriptaddress,notitlepage,nofootinbib,preprintnumber,twocolumn]
{revtex4-1}
\usepackage{amsfonts,amssymb,amsmath,graphicx,color,bm,enumitem}
\usepackage[utf8]{inputenc}
\usepackage{amsthm}
\usepackage[english]{babel}
\definecolor{ultramarine}{rgb}{0.07, 0.04, 0.56}
\definecolor{cadmiumgreen}{rgb}{0.0, 0.42, 0.24}
\definecolor{indigo(dye)}{rgb}{0.0, 0.25, 0.42}
\usepackage[linktocpage=true]{hyperref}
\hypersetup{
colorlinks=true,
citecolor=blue,
linkcolor=green,
urlcolor=indigo(dye),
}

\newcommand{\be}{\begin{eqnarray}}  
\newcommand{\ee}{\end{eqnarray}}
\newcommand{\bem}{\begin{pmatrix}}
\newcommand{\eem}{\end{pmatrix}}

\newcommand{\E}{\mathcal{E}}
\newcommand{\F}{\mathcal{F}}
\newcommand{\G}{\mathcal{G}}

\newcommand{\Q}{\mathcal{Q}}
\newcommand{\R}{\mathcal{R}}
\newcommand{\V}{\mathcal{V}}
\newcommand{\W}{\mathcal{W}}
\newcommand{\D}{\mathcal{D}}
\newcommand{\X}{\mathcal{X}}
\newcommand{\Y}{\mathcal{Y}}
\newcommand{\C}{\mathcal{C}}
\newcommand{\B}{\mathcal{B}}
\newcommand{\tr}{\tilde{r}}

\begin{document}

\title{
Disformal transformation of physical quantities associated with relativistic stars
}

\author{Masato Minamitsuji}
\affiliation{Centro de Astrof\'{\i}sica e Gravita\c c\~ao  - CENTRA, Departamento de F\'{\i}sica, Instituto Superior T\'ecnico - IST, Universidade de Lisboa - UL, Av. Rovisco Pais 1, 1049-001 Lisboa, Portugal}

\begin{abstract}
We investigate how physical quantities associated with relativistic stars in the Jordan and Einstein frames are related by the generalized disformal transformations constructed by the scalar and vector fields within the slow-rotation approximation. We consider the most general scalar disformal transformation constructed by the scalar field, and by the vector field without and with the $U(1)$ gauge symmetry, respectively. At the zeroth order of the slow-rotation approximation, by imposing that both the metrics of the Jordan and Einstein frames are asymptotically flat, we show that the Arnowitt-Deser-Misner mass is frame-invariant. At the first order of the slow-rotation approximation, we discuss the disformal transformations of the frame-dragging function, angular velocity, angular momentum, and moment of inertia of the star. We show that the angular velocity of the star is frame-invariant in all the cases. While the angular momentum and moment of inertia are invariant under the scalar disformal transformation, they are not under the vector disformal transformation without and with the $U(1)$ gauge symmetry. 
\end{abstract}
\pacs{04.50.-h, 04.50.Kd, 98.80.-k}
\maketitle

\section{Introduction}
\label{sec1}

While general relativity (GR) has passed all the experimental tests
in weak gravity regimes~\cite{Will:2014kxa},
with the latest measurements of gravitational waves~\cite{Berti:2015itd,Berti:2018vdi,Berti:2018cxi}
and black hole shadows \cite{Psaltis:2020lvx,Psaltis:2018xkc},
a new frontier for testing GR in strong-field regimes has opened.
Black holes 
are the most fundamental objects in theories of gravitation.
While many modified theories of gravitation 
share the same black hole solutions with GR
\cite{Psaltis:2007cw,Motohashi:2018wdq},
other theories also admit black solutions 
with the nontrivial profiles of the non-metric field.

Relativistic stars as neutron and quark stars \cite{Lattimer:2006xb,Ozel:2016oaf} 
are known to be the second most compact objects in the Universe.
Scalar-tensor theories of gravitation
with nonminimal coupling to the matter sector
admit large deviations from GR
through the mechanism called
{\it spontaneous scalarization}~\cite{Damour:1993hw,Damour:1996ke,Harada:1997mr,Harada:1998ge,Novak:1998rk,Palenzuela:2013hsa,Sampson:2014qqa,Pani:2014jra,Silva:2014fca},
triggered by the conformal coupling
to the matter sector
\be
\label{conformal}
{\tilde g}_{\mu\nu}= \C (\phi) g_{\mu\nu},
\ee
where
the Greek induces $(\mu,\nu,\cdots)$ run the 
four-dimensional spacetime
and $\C(\phi)\left(>0\right)$
is the regular function of
the scalar field $\phi$.
In this paper, 
we call the frames with the metrics ${g}_{\mu\nu}$ and ${\tilde g}_{\mu\nu}$
the {\it Einstein} and {\it Jordan} frames,
respectively,
where the action of the matter sector $\Psi_m$ is given by
\be
\label{matter}
S\supset 
S_m\left[
{\tilde g}_{\mu\nu},
\Psi_m
\right]:=
\int d^4x \sqrt{-{\tilde g}}
{\cal L}_m
\left[
{\tilde g}_{\mu\nu},
\Psi_m
\right].
\ee
The energy-momentum tensors of matter
in the Jordan and Einstein frames,
${\tilde T}^{\mu\nu}_{(m)}$
and 
$T^{\mu\nu}_{(m)}$,
respectively,
are related by
\begin{eqnarray}
\label{rel}
T_{(m)}^{\mu\nu}
&:=&
\frac{2}{\sqrt{-g}}
\frac{\delta\left(\sqrt{- {\tilde g}} {\cal L}_m\right)}
       {\delta g_{\mu\nu}}
=\sqrt{\frac{\tilde g}{g}}
\frac{\delta {\tilde g}_{\alpha\beta}}
       {\delta {g}_{\mu\nu}}
{\tilde T}_{(m)}^{\alpha\beta},
\end{eqnarray}
with 
\be
{\tilde T}_{(m)}^{\mu\nu}
&:=&
\frac{2}{\sqrt{-{\tilde g}}}
\frac{\delta\left(\sqrt{- {\tilde g}} {\cal L}_m\right)}
       {\delta {\tilde g}_{\mu\nu}}.
\ee
Ref.~\cite{Minamitsuji:2016hkk} considered
spontaneous scalarization in the model with the scalar disformal coupling
\begin{eqnarray}
\label{disformal0}
{\tilde g}_{\mu\nu}
=
\C(\phi)
\left(
    g_{\mu\nu}
+ \B(\phi)\phi_\mu \phi_\nu
\right),
\end{eqnarray}
where
$\B(\phi)$ is the regular function of $\phi$,
$\phi_\mu:=\nabla_\mu \phi$
represents the covariant derivative of the scalar field,
and
the canonical kinetic term of the scalar field
is defined by 
\be
\label{canonical}
\X:=-\frac{1}{2}g^{\mu\nu}\phi_\mu\phi_\nu.
\ee

The extension of spontaneous scalarization 
to the vector field $A_\mu$,
i.e., 
{\it spontaneous vectorization},
has been investigated in Refs.
\cite{Ramazanoglu:2017xbl,Ramazanoglu:2019gbz,Annulli:2019fzq,Ramazanoglu:2019jrr,Kase:2020yhw,Minamitsuji:2020pak}.
Ref.~\cite{Annulli:2019fzq}
considered a model with nonminimal coupling to the Ricci tensor $R^{\mu\nu}A_\mu A_\nu$
and the Ricci scalar $R g^{\mu\nu} A_\mu A_\nu$,
and Ref.~\cite{Kase:2020yhw} studied
a model
with nonminimal coupling to the Einstein tensor $G^{\mu\nu}A_\mu A_\nu$
in the generalized Proca theories \cite{Heisenberg:2014rta,DeFelice:2016cri}.
Ref. \cite{Ramazanoglu:2017xbl}
considered the vector conformal coupling 
\begin{eqnarray}
\label{disformalv}
{\tilde g}_{\mu\nu}
=
\D(\Y)    g_{\mu\nu},
\end{eqnarray}
where $\D(\Y)$ is the regular function of 
the spacetime norm of the vector field,
\be
\label{defy}
\Y:=-\frac{1}{2}g^{\mu\nu}A_\mu A_\nu.
\ee
Ref. \cite{,Ramazanoglu:2019jrr,Minamitsuji:2020pak}
argued the case of the vector disformal coupling
\be
\label{pd}
{\tilde g}_{\mu\nu}= g_{\mu\nu}+ {\E}(\Y)A_\mu A_\nu,
\ee
where $\E(\Y)$ is the regular function of $\Y$.

However, 
before exploring more generalized models 
of spontaneous scalarization and vectorization,
as well as 
other models
such as {\it spontaneous spinorization} \cite{Ramazanoglu:2018hwk,Minamitsuji:2020pak},
we have to clarify the dependence of physical quantities 
associated with relativistic stars
on the choice of the frames.
Although the physical frame in these models
is the Jordan frame
where the scalar or vector field is not directly coupled to the matter sector,
in many cases
it is more convenient to evaluate the same quantities in the Einstein frame.
In such a case,
in order to translate the quantities defined in the Einstein frame
to those in the Jordan frame, 
we have to clarify
the relation of them defined in both the frames
beforehand.
Since the frame transformation mathematically 
corresponds to the change of the units or the measures,
we naively expect that the observationally relevant quantities
should be independent of the choice of the frames.
However, 
the frame dependence of
physical quantities 
is rather a subtle issue
and 
should be clarified in each of the different cases.  
For instance, 
in the context of cosmology in scalar-tensor theories, 
the frame-invariance
of  
cosmological redshift,
distance-duality relation,
scalar and tensor cosmological perturbations,
anisotropy of Cosmic Microwave Backgrounds,
and so on
have been argued
under the conformal and disformal transformations
(see e.g., Refs.~
\cite{Makino:1991sg,Gong:2011qe,Chiba:2008ia,Chiba:2013mha,Minamitsuji:2014waa,Tsujikawa:2014uza,Watanabe:2015uqa,Motohashi:2015pra,Chiba:2020mte}).

In the case of the systems associated with relativistic stars, 
to our knowledge,
the frame-invariance of the observables
has been less studied.
In the case of scalar-tensor theories
with the disformal coupling \eqref{disformal0}
including the ordinary conformal coupling case \eqref{conformal}.
Ref.~\cite{Minamitsuji:2016hkk}
showed
that 
the Arnowitt-Deser-Misner (ADM) mass,
frame-dragging function,
angular velocity,
angular momentum 
and 
moment of inertia of the star
are frame-invariant 
with the context of the slow-rotation approximation
\cite{Hartle:1967he,Hartle:1968si}.

The purpose of this paper
is 
to extend the previous work \cite{Minamitsuji:2016hkk}
to more general classes 
of the disformal transformation
constructed by the scalar and vector fields.
More specifically,
we will focus on the following three classes
of the disformal transformations;

\begin{description}

\item[(1)]
most general disformal transformation
constructed by the scalar field and its first-order derivative;
\begin{eqnarray}
\label{disformal}
{\tilde g}_{\mu\nu}
=
\C(\X,\phi)
\left(
    g_{\mu\nu}
+ \B(\X,\phi)\phi_\mu \phi_\nu
\right),
\end{eqnarray}
where
$\C$ and $\B$ are the regular functions of  $\X$ and $\phi$ 
defined in Eq. \eqref{canonical} (see Sec. \ref{sec3}).

\item[(2)] 
disformal transformation constructed by the vector field
without the $U(1)$ gauge symmetry;
\begin{eqnarray}
\label{disformalv2}
{\tilde g}_{\mu\nu}
=
\D(\Y)  
\left(  
g_{\mu\nu}
+
\E(\Y)
A_\mu A_\nu
\right),
\end{eqnarray}
which generalizes the pure conformal 
and disformal transformations 
\eqref{disformalv} and \eqref{pd}
(see Sec. \ref{sec4}).

\item[(3)] 
disformal transformation constructed by the vector field
with the $U(1)$ gauge symmetry;
\begin{eqnarray}
\label{disformalf2}
{\tilde g}_{\mu\nu}
=
\V(\F,\G)
\left(
  g_{\mu\nu}
+\W(\F,\G)  
  g^{\alpha\beta}
F_{\mu\alpha} F_{\nu\beta}  
\right),
\end{eqnarray}
where
$\V$ and $\W$ are the regular functions of
\begin{subequations}
\begin{eqnarray}
\F
&:=&g^{\mu\alpha}g^{\nu\beta} F_{\mu\nu}F_{\alpha\beta},
\\
\G
&:=& 
g^{\mu\nu}
g^{\alpha\beta}
g^{\rho\sigma}
g^{\gamma\delta}
F_{\mu\alpha}
F_{\rho\beta} 
F_{\sigma\gamma} 
F_{\nu\delta},
\end{eqnarray}
\end{subequations}
with the $U(1)$ gauge field strength
\be
F_{\mu\nu}:=\partial_\mu A_\nu-\partial_\nu A_\mu,
\ee
(see Sec. \ref{sec5}).

\end{description}
The invertible scalar disformal transformation \eqref{disformal} 
can frame 
the degenerate higher-order scalar-tensor
theories \cite{Langlois:2015cwa,Achour:2016rkg,BenAchour:2016fzp,Langlois:2018dxi},
which are known as the most general single-field
scalar-tensor theories without Ostrogradsky instabilities~\cite{Woodard:2015zca}.
The invertible vector disformal transformation 
without the $U(1)$ gauge symmetry \eqref{disformalv2}
can map the generalized Proca theories
\cite{Heisenberg:2014rta,Tasinato:2014eka,DeFelice:2016cri},
which are known as the most general vector-tensor theories
with the second-order equations of motion,
to the extended vector-tensor theories 
without Ostrogradsky instabilities
constructed in Ref.~\cite{Kimura:2016rzw}.
Similarly,
the invertible vector disformal transformation \eqref{disformalf2}
constitutes
the most general frame transformation
with the vector field
under the $U(1)$ gauge symmetry,
which maps the Einstein-Maxwell theory
to the degenerate higher-order vector-tensor theories
with the $U(1)$ gauge symmetry
and 
without Ostrogradsky instabilities~\cite{Gumrukcuoglu:2019ebp,DeFelice:2019hxb}.

This paper is constructed as follows:
In Sec. \ref{sec2},
we introduce the slowly rotating star supported by the perfect fluid
and
the spacetime metric in the slow-rotation approximation.
In Sec. \ref{sec3}, 
we discuss
how physical quantities associated with relativistic stars 
are transformed
under the disformal transformation
constructed by the scalar field \eqref{disformal}.
In Secs. \ref{sec4} and \ref{sec5},
we study the same subject
in the case of the disformal transformations constructed by the vector field 
without and with the $U(1)$ symmetry,
Eqs. \eqref{disformalv2} and \eqref{disformalf2},
respectively.
The last Sec. \ref{sec6}
is devoted to 
giving a brief summary and conclusion.

\section{Slowly rotating stars}
\label{sec2}

We consider the rigidly rotating star with the constant 
angular velocity $\Omega$,
and adopt the slow-rotation approximation \cite{Hartle:1967he,Hartle:1968si},
where 
the variables of the metric and other degrees of freedom
are expanded 
with respect to the small parameter $\epsilon (\ll 1)$,
which is of  the order of the dimensionless angular speed of the star
$ {\Omega}\tilde{\R} (\ll 1)$, 
with $\tilde{\cal R}$ being the radius of the star.
At ${\cal O} (\epsilon^0)$,
the metric and the scalar or vector field equations of motion
coincide
with those in the static and spherically symmetric spacetime.
At ${\cal O} (\epsilon^1)$,
the leading-order corrections due to the slow rotation
of the star,
which appear
in the frame-dragging component of the spacetime metric $g_{t\varphi}$
and  
the azimuthal component of the vector field $A_\varphi$,
are obtained
on top of the static and spherically symmetric backgrounds.

Up to ${\cal O} (\epsilon^1)$,
the metric inside and outside of the star
can be expressed as  
\be
\label{ansatza}
g_{\mu\nu}
  dx^\mu dx^\nu
&=&
-f(r)dt^2
+\frac{dr^2}{h(r)}
+
r^2 
\left(
  d\theta^2
+\sin^2\theta 
  d\varphi^2
\right)
\nonumber\\
&+&
2
\epsilon
r^2
\left(
  \omega(r)
-\Omega
\right)
\sin^2\theta
dt d\varphi
+
{\cal O}
(\epsilon^2),
\ee
where $t$, $r$, and $(\theta,\varphi)$
represent the time, radial, and angular coordinates, respectively,
$f(r)$ and $h(r)$ are the functions of $r$,
and 
$\omega(r)$ and $\Omega$ represent
the frame-dragging function and the angular velocity of the star.
$f(r)$, $h(r)$, and $\omega(r)$
are continuous across the surface of the star.

We impose
the asymptotic flatness of the Einstein-frame metric $g_{\mu\nu}$
in the limit of $r\to \infty$,
which is given by
\begin{subequations}
\label{asym}
\be
&&
f(r\to\infty)
\to 
{\rm const},
\\
&&
h(r\to \infty)
\to 
1,
\\
&&
\omega(r\to \infty)
\to
\Omega.
\ee
\end{subequations}
We assume that 
the Jordan-frame metric
${\tilde g}_{\mu\nu}$, Eq. \eqref{ansatzb},
satisfies the {\it circularity conditions}~\cite{circ1,circ2,waldbook};
if the two commuting Killing vector fields
in the stationary and axisymmetric spacetime
$\xi^\mu=(\partial/\partial t)^\mu$
and 
$\sigma^\mu=(\partial/\partial \varphi)^\mu$
satisfy
\begin{eqnarray}
\label{circularity2}
\sigma_{[\mu} \xi_{\nu}{\tilde \nabla}_{\alpha} \xi_{\beta]}=0,
\qquad 
\sigma_{[\mu} \xi_{\nu}{\tilde \nabla}_{\alpha} \sigma_{\beta]}=0,
\end{eqnarray}
where ${\tilde \nabla}_\mu$ represents
the covariant derivative associated with the metric ${\tilde g}_{\mu\nu}$,
then the spacetime metric reduces to the block-diagonal form 
with ${\tilde g}_{t{\tilde r}}={\tilde g}_{t\theta}={\tilde g}_{{\tilde r}\varphi}={\tilde g}_{\theta\varphi}=0$,
which in the slow-rotation approximation reduces to the form \cite{Minamitsuji:2020jvf};
\begin{eqnarray}
\label{ansatzb}
{\tilde g}_{\mu\nu}dx^\mu dx^\nu
&=&
-{\tilde f} (\tr)
dt^2
+
\frac{d{\tilde r}^2}{{\tilde h}(\tr)}
+
{\tilde r}^2
\left(
d\theta^2
+
\sin^2\theta 
d\varphi^2
\right)
\nonumber\\
&+&
2\epsilon
{\tilde r}^2 
\sin^2\theta
\left(
\tilde{\omega}(\tr)
-\tilde{\Omega}
\right)
dt d\varphi
\nonumber\\
&+&
{\cal O} (\epsilon^2),
\end{eqnarray}
where the new radial coordinate ${\tilde r}(r)$,
and the new metric functions ${\tilde f} ({\tilde r})$, ${\tilde h}({\tilde r})$,
and ${\tilde \omega}({\tilde r})$
are related to the original ones by the direct comparison
with Eqs. \eqref{disformal}, \eqref{disformalv2}, and \eqref{disformalf2}.
This requirement is automatically satisfied for the disformal transformations 
\eqref{disformal} and \eqref{disformalf2},
but restricts the form of the vector field
in the case of the vector disformal transformation 
without the $U(1)$ gauge symmetry \eqref{disformalv2}
(see Sec. \ref{sec4}).
We assume the circularity conditions \eqref{circularity2} for several reasons.
First,
in order for the physical quantities to be frame-invariant,
we expect that the metrics 
in the Jordan and Einstein frames
should share the same form,
and that there always exist counterparts of the metric and matter variables
in both the frames. 
Second, 
imposing the circularity conditions \eqref{circularity2}
simplifies the computations.
In the case of  black holes,
it was recently argued 
in Ref. \cite{Xie:2021bur}
that in the effective theory description of
non-GR gravitational theories 
the spacetime of stationary,
axisymmetric,
and asymptotically flat solutions
has to be circular,
when the solutions are constructed 
perturbatively
from the GR solutions.
The similar properties may exist also for 
the solutions of relativistic stars
constructed perturbatively from the GR ones.

On the other hand, 
the physical consequences of the violation of the circularity conditions
on the dynamical quantities and their frame transformations
are worth being investigated
and for now will be left for future work.
We expect
that
no physical pathology associated with
the violation of the circularity conditions
would arise,
such as instabilities.
However, 
the noncircular spacetime around the star
would modify the motion of photons and particles
in the vicinity of the stars,
and
affect the measurements 
of pulsars and gravitational waves.
Thus, 
if no peculiar deviation is detected in the future,
the deviations from the circular spacetimes would
be significantly constrained. 

Here, 
we require that
${\tilde r}(r)$ is the monotonically increasing function of $r$,
and
impose the asymptotic flatness of ${\tilde g}_{\mu\nu}$
in the limit of ${\tilde r}\to \infty$
\begin{subequations}
\label{asym2}
\be
&&
{\tilde f}({\tilde r}\to\infty)
\to 
{\rm const}>0,
\\
&&
{\tilde h}({\tilde r}\to \infty)
\to 
1,
\\
&&
{\tilde \omega}({\tilde r}\to \infty)
\to
{\tilde  \Omega}.
\ee
\end{subequations}


We consider the star composed of matter
whose energy-momentum tensor in the Einstein frame 
is given in the perfect fluid form
\begin{eqnarray}
\label{em1}
T_{(m)\mu\nu}
&=&
{\rho} u_\mu u_\nu
+{p}_r k_\mu k_\nu
\nonumber\\
&+&
{p}_t
\left(
  {g}_{\mu\nu}
+{u}_\mu  {u}_\nu
-{k}_\mu  {k}_\nu
\right),
\end{eqnarray}
where the components of the vector fields,
up to ${\cal O}(\epsilon^1)$,
are defined by 
\begin{subequations}
\label{em2}
\begin{eqnarray}
u_\mu
&=&
\left(
-\sqrt{-g_{tt}},0,0,
\epsilon \frac{r^2 \omega\sin^2\theta}{\sqrt{-g_{tt}}}
\right)
+{\cal O} (\epsilon^2),
\\
k_\mu
&=&
\left(
0,\sqrt{g_{rr}},0,0
\right)
+{\cal O} (\epsilon^2).
\end{eqnarray}
\end{subequations}
Correspondingly,
the energy-momentum tensor of matter 
in the Jordan frame is also assumed 
to be given by the perfect fluid form
Eqs. \eqref{em1}-\eqref{em2}
with the replacement of all the quantities to those with `tilde',
which are related  
through Eq. \eqref{rel}
with Eqs. 
\eqref{disformal},
\eqref{disformalv2}, 
and
\eqref{disformalf2}.
The surface of the star is fixed by the condition ${\tilde p}_r({\tilde\R})=0$.
As we will see in Secs. \ref{sec3}-\ref{sec5} ,
in general
the condition ${\tilde p}_r(\tilde{\R})=0$
does not imply that $p_r (\R)=0$ with $\tilde{\R}={\tilde \R}\left({\R}\right)$,
and 
the concrete study in each case
of the disformal transformations,
Eqs. \eqref{disformal}, \eqref{disformalv2}, and \eqref{disformalf2},
is necessary.

At ${\cal O} (\epsilon^0)$,
the mass functions in both frames,
$M(r)$ and ${\tilde M} ({\tilde r})$,
are defined by
\begin{subequations}
\label{adms}
\be
M(r)
&=&\frac{r}{2}
\left(
1-h(r)
\right),
\\
{\tilde M}
({\tilde r})
&=&
\frac{{\tilde r}}{2}
\left(
1
-{\tilde h} ({\tilde r})
\right),
\ee
\end{subequations}
whose limits to $r\to \infty$ and ${\tilde r}\to \infty$
provide the ADM masses,
respectively,
\begin{subequations}
\label{adms2}
\be
&&
M_{\rm ADM}:=M(r\to\infty),
\\
&&
{\tilde M}_{\rm ADM}
:={\tilde M} ({\tilde r}\to \infty).
\ee
\end{subequations}
Up to ${\cal O} (\epsilon^1)$,
the angular momenta and moments of inertia 
in both frames
are given, 
respectively,
by
\begin{subequations}   
\label{am}
\begin{eqnarray}
{\cal J}
&:=&
\int dr d\theta d\varphi
r^2\sin\theta 
\sqrt{\frac{f}{h}}
T_{(m)}{}^t{}_\varphi,
\\
{\tilde {\cal J}}
&:=&
\int d{\tilde r} d\theta d\varphi
{\tilde r}^2\sin\theta 
\sqrt{\frac{\tilde f}{\tilde h}}
{\tilde T}_{(m)}{}^t{}_\varphi,
\end{eqnarray}
\end{subequations}
and 
\begin{subequations}
\label{moi}
\begin{eqnarray}
{\cal I}:=\frac{{\cal J}}
                  {\Omega},
\\
{\tilde {\cal I}}:
=\frac{{\tilde {\cal J}}}
                  {{\tilde \Omega}}.
\end{eqnarray}
\end{subequations}

\section{The case of the scalar disformal transformation}
\label{sec3}

In this section, 
we consider the scalar-tensor theories
\be
S=S_g \left[g_{\mu\nu},\phi\right]
+  
S_m\left[{\tilde g}_{\mu\nu},\Psi_m\right],
\ee
where the matter action is given by Eq. \eqref{matter}
and 
the relation between the two metrics is given 
by Eq. \eqref{disformal}.

In the slow-rotation approximation,
we assume the spacetime metric~\eqref{ansatza}
and the scalar field given by
\begin{subequations}
\begin{eqnarray}
\phi
&=&\phi_0(r)
+{\cal O} (\epsilon^2),
\\
\X
&=&
\X_0(r)
+{\cal O} (\epsilon^2)
:=
-\frac{h(r)\phi_0'(r)^2}{2}
+{\cal O} (\epsilon^2).
\end{eqnarray}
\end{subequations}
The functions $\C$ and $\B$
can be expanded as 
\begin{subequations}
\be
\C
&=&
\C_0(r)
+{\cal  O}(\epsilon^2)
\nonumber\\
&:=&
\C
\left(
\X_0(r),\phi_0(r)
\right)
+{\cal  O}(\epsilon^2),
\\
\B
&=&
\B_0(r)
+{\cal  O}(\epsilon^2)
\nonumber\\
&:=&
\B
\left(
\X_0(r),\phi_0(r)
\right)
+{\cal  O}(\epsilon^2),
\\
\C_{0\X}(r)&:=&\C_{,\X}(\X_0(r),\phi_0(r)),
\\ 
\B_{0\X}(r)&:=&\B_{,\X}(\X_0(r),\phi_0(r)),
\ee
\end{subequations}
where
$\C_{,\X}:= \partial \C/\partial \X$
and 
$\B_{,\X}:= \partial \B/\partial \X$.
Comparing frames relates
the radial coordinates and the metric components
in the Jordan and Einstein frames,
respectively,
as
\be
{\tilde r}:=\sqrt{\C_0(r)}r,
\ee
and 
\begin{subequations}
\label{tf1}
\begin{eqnarray}
{\tilde f}
({\tilde r})
&=&
\C_0 (r)
f(r),
\\
{\tilde h}
({\tilde r})
&=&
\frac{
h(r)}
{1
+\B_0(r) h(r) \phi_0'(r)^2}
\nonumber\\
&\times&
\left(1+\frac{r}{2\C_0(r)}\frac{d\C_0(r)}{dr}\right)^2,
\\
{\tilde  \omega}
({\tilde r})
-
{\tilde \Omega}
&=&
 \omega(r)
-\Omega,
\end{eqnarray}
\end{subequations}
with 
\be
\frac{d\C_0(r)}{dr}
=-
\frac{\C_{0\X}(r) \phi_0'(r)}{2}
 \left(h'(r)\phi'_0(r) +2h(r)\phi_0''(r)\right).
\nonumber\\
\ee
In order to be compatible with the asymptotic flatness
in the Jordan frame \eqref{asym2},
we impose Eq. \eqref{asym} and 
\begin{subequations}
\label{af1}
\begin{eqnarray}
\C_0(r)
&\to& {\rm const}>0,
\\
 \phi_0'(r)^2
\times \B_{0}(r)
&\to& 0,
\\
rh'(r)\left(\phi_0'(r)^2\right)
\times \C_{0\X}(r)
&\to & 0,
\\
r\frac{d}{dr}\left(\phi_0'(r)^2\right) 
\times \C_{0\X}(r)
&\to & 0,
\end{eqnarray}
\end{subequations}
as $r\to \infty$.
Under the conditions \eqref{af1},
the ADM masses \eqref{adms2} 
are shown to be invariant,
\begin{eqnarray}
{\tilde M}_{\rm ADM}
=
M_{\rm ADM}.
\end{eqnarray}
Eq. \eqref{rel} reduces to 
\begin{eqnarray}
T_{(m)}^{\mu\nu}
&=&
\C^3 \sqrt{1-2\B\X}
\left[
{\tilde T}_{(m)}^{\mu\nu}
+
\frac{1}{2}
  g^{\mu \alpha }g^{\nu\beta} 
  \phi_\alpha\phi_\beta 
\right.
\nonumber\\
&&
\left.
\times 
  \left(
\B_{,\X}
{\tilde T}_{(m)}^{\rho\sigma} 
\phi_\rho  \phi_\sigma
+ \frac{\C_{,\X}}{\C^2}
{\tilde T}_{(m)}
\right)
\right],
\end{eqnarray}
where we have defined  
${\tilde T_{(m)}}:={\tilde g}_{\alpha\beta} {\tilde T}_{(m)}^{\alpha\beta}$.
At ${\cal O}(\epsilon^0)$,
namely 
the energy density and pressures in both the frames
are related
\begin{subequations}
\begin{eqnarray}
\rho(r)
&=&
\C_0(r)^2\sqrt{1+\B_0(r) h(r) \phi'(r)^2}
{\tilde \rho}({\tilde r}),
\\
p_r(r)
&=&
\frac{\C_0(r)^2}{\sqrt{1+\B_0(r) h(r) \phi'(r)^2}}
\nonumber\\
&\times &
\left[
{\tilde p}_r({\tilde r})
+\frac{h(r) \left(\phi'(r)\right)^2}{2}
\left(
\B_{0\X}(r) h(r){\tilde p}_r ({\tilde r})\left(\phi'(r)\right)^2
\right.
\right.
\nonumber\\
&&
\left.
\left.
+\frac{\C_{0\X}(r) {\tilde T}_{(m,0)} ({\tilde r})}{\C_0(r)}
\left(
1+\B_0(r) h(r) \phi'(r)^2
\right)
\right)
\right],
\nonumber
\\
\\
p_t(r)
&=&
\C_0(r)^2\sqrt{1+\B_0(r) h(r) \phi'(r)^2}
{\tilde p}_t({\tilde r}),
\end{eqnarray}
\end{subequations}
where
${\tilde T}_{(m,0)}({\tilde r})=-{\tilde\rho}({\tilde r})+{\tilde p}_r({\tilde r})+2{\tilde p}_t({\tilde r})$,
and 
at ${\cal O}(\epsilon^1)$,
\begin{subequations}
\begin{eqnarray}
{\tilde  \omega}
({\tilde r})
&=&
{\omega}(r),
\\
{\tilde  \Omega}
&=&
{\Omega}.
\end{eqnarray}
\end{subequations} 
In general,
the isotropic fluid in the Jordan frame ${\tilde p}_r({\tilde r})={\tilde p}_t({\tilde r})$
is mapped to the anisotropic fluid $p_r(r)\neq p_t(r)$.
In general, 
the surface condition ${\tilde p}_r(\tilde{\R})=0$
in the Jordan frame
also corresponds to the condition $p_r (\R)=0$
in the Einstein frame with $\tilde{\R}={\tilde \R}\left({\R}\right)$
for $\C_{0\X}(\R)=0$.
The frame-dragging function and the angular velocity of the star
are disformally invariant.

It is straightforward
to confirm
that 
up to ${\cal O}\left(\epsilon^0\right)$
the angular momenta and moments of inertia,
Eqs. \eqref{am} and \eqref{moi},
respectively,
are also disformally invariant
\begin{subequations} 
\begin{eqnarray}
{\tilde {\cal J}}&=&{\cal J},
\\
{\tilde {\cal I}}&=&{\cal I}.
\end{eqnarray}
\end{subequations} 
Our results correspond to the extension of 
the work \cite{Minamitsuji:2016hkk}
for the subclass of Eqs. $\C=\C(\phi)$ and $\B=\B(\phi)$,
which can frame the Horndeski theories
\cite{Horndeski:1974wa,Nicolis:2008in,Deffayet:2009mn,Kobayashi:2011nu}.

\section{The vector disformal transformation without the $U(1)$ gauge symmetry}
\label{sec4}

In this section, 
we consider the vector-tensor theories
without the $U(1)$ gauge symmetry
\be
S=S_g \left[g_{\mu\nu},A_\mu\right]
+  S_m\left[{\tilde g}_{\mu\nu},\Psi_m\right],
\ee
where the matter action is given by Eq. \eqref{matter}
and 
the relation between the two metrics is given 
by Eq.~\eqref{disformalv2}.
Eq. \eqref{rel} reduces to 
\begin{eqnarray}
\label{interim}
T_{(m)}^{\mu\nu}
&=&
\D(\Y)^3
\sqrt{1-2\E(\Y) \Y}
\left[
{\bar T}_{(m)}^{\mu\nu}
\right.
\nonumber\\
&&
+
\left.
g^{\mu\alpha}
g^{\nu\beta}
A_\alpha A_\beta
\Q
\right],
\end{eqnarray}
where
we have defined
\be
\label{calq}
\Q:= 
\frac{\D_{,\Y}(\Y)}{2\D(\Y)^2}
{\tilde T}_{(m)}
+
\frac{\E_{,\Y}(\Y)}{2}
{\tilde T}_{(m)}^{\rho\sigma}
A_\rho A_\sigma,
\ee
with 
$\D_{,\Y}:= \partial \D(Y)/\partial \Y$
and 
$\E_{,\Y}:= \partial \E(Y)/\partial \Y$.
In the slow-rotation approximation,
the vector field should have the following form \cite{Minamitsuji:2020jvf}
\begin{eqnarray}
\label{vectorfield}
A_\mu dx^\mu
&=&
  A_t (r)dt
+A_r (r) dr
+
\epsilon
r^2 a_3(r)\sin^2\theta 
d\varphi
\nonumber\\
&+&
{\cal O}
\left(\epsilon^2\right).
\end{eqnarray}
In general, the metric \eqref{disformalf2} 
contains the components
\begin{subequations}
\be
{\tilde g}_{tr}
&=&\D_{0}(r) \E_0(r) A_t(r) A_r(r),
\\
{\tilde g}_{r\varphi}
&=&
\epsilon
r^2 a_3(r) A_r(r)  \D_0(r) \E_0(r)\sin^2\theta,
\ee
\end{subequations}
and ${\tilde g}_{t\theta}={\tilde g}_{\theta\varphi}=0$,
where we have defined
\begin{subequations}
\be
\D(Y)
&=&
\D_0(r)
+{\cal O} (\epsilon^2)
\nonumber\\
&:=&
\D
\left(\frac{1}{2f(r)}A_t(r)^2 -\frac{h(r)}{2}A_r(r)^2\right)
+
{\cal O} (\epsilon^2),
\nonumber
\\
\\
\D_{0\Y}(r)
&:= &
\D_{,\Y} \left(\frac{1}{2f(r)}A_t(r)^2 -\frac{h(r)}{2}A_r(r)^2\right),
\\
\E(Y)
&=&
\E_0(r)
+{\cal O} (\epsilon^2)
\nonumber\\
&:=&
\E
\left(\frac{1}{2f(r)}A_t(r)^2 -\frac{h(r)}{2}A_r(r)^2\right)
+
{\cal O} (\epsilon^2),
\nonumber
\\
\\
\E_{0\Y}(r)
&:= &
\E_{,\Y} \left(\frac{1}{2f(r)}A_t(r)^2 -\frac{h(r)}{2}A_r(r)^2\right).
\ee
\end{subequations}
In order to be compatible with the circularity conditions \eqref{circularity2}
as well as the nonzero $\D_0(r)\neq 0$,
we impose either of the following two conditions separately:

\begin{itemize}

\item
the pure conformal coupling \eqref{disformalv};
\be
\label{pure_conf}
\E(\Y)=0.
\ee

\item
the vanishing radial component of the vector field
\be
\label{ar0}
A_r(r)=0.
\ee

\end{itemize}
Relativistic star solutions with
nontrivial $A_t(r)$ and Eq. \eqref{ar0}
have been investigated
in the models in the vector-tensor theories
\cite{Ramazanoglu:2017xbl,Ramazanoglu:2019gbz,Annulli:2019fzq,Ramazanoglu:2019jrr,Kase:2020yhw,Minamitsuji:2020pak,Kase:2017egk}.

\subsection{The pure conformal case \eqref{pure_conf}}
\label{sec41}

First, 
we consider the case of Eq.~\eqref{disformalv}
which relates the radial coordinates and metric components
in the Jordan and Einstein frames,
respectively, as 
\begin{eqnarray}
\label{tr2}
{\tilde r}:=\sqrt{\D_0(r)}r,
\end{eqnarray}
and
\begin{subequations}
\label{tf2}
\begin{eqnarray}
{\tilde f}
({\tilde r})
&=&
\D_0(r) f(r),
\\
{\tilde h} ({\tilde r})
&=&
h(r) \left(1+\frac{r}{2\D_0(r)}\frac{d\D_0}{dr}(r)\right)^2,
\\
{\tilde  \omega} ({\tilde r})
-
{\tilde \Omega}
&=&
 \omega(r)
-\Omega,
\end{eqnarray}
\end{subequations}
where
\begin{eqnarray}
\label{tf22}
\frac{d\D_0}{dr}(r)
&=&
\D_{0\Y}(r)
\left[
\frac{A_t(r)}{2f(r)^2}
\left(
2f (r)A_t'(r)
-A_t(r)f'(r)
\right)
\right.
\nonumber\\
&&
\left.
-
\frac{A_r(r)}{2}
\left(
h'(r) A_r(r)+ 2A_r'(r) h(r)
\right)
\right].
\end{eqnarray}
In order to be compatible with the asymptotic flatness
in the Jordan frame \eqref{asym2},
we impose Eq. \eqref{asym} and 
\begin{subequations}
\label{af2}
\begin{eqnarray}
\D_0(r)
&\to& 
{\rm const}>0,
\\
r
\frac{d}{dr}
\left(A_t(r)^2\right)
\times \D_{0\Y}(r)
&\to & 
0,
\\
r
f'(r)
A_t(r)^2
\times 
\D_{0\Y}(r)
&\to & 
0,
\\
r
\frac{d}{dr}
\left(A_r(r)^2\right)
\times \D_{0\Y}(r)
&\to & 
0,
\\
r
h'(r)
A_r(r)^2
\times \D_{0\Y}(r)
&\to & 
0,
\end{eqnarray}
\end{subequations}
as $r\to \infty$.
Under the conditions \eqref{af2},
the ADM masses \eqref{adms2} 
are shown to be invariant
\begin{eqnarray}
{\tilde M}_{\rm ADM}
=
M_{\rm ADM}.
\end{eqnarray}
The energy-momentum tensors of matter
are related via Eq. \eqref{interim} 
with Eqs. \eqref{calq} and \eqref{pure_conf}.
Assuming the perfect fluid form \eqref{em1} and \eqref{em2},
respectively,
comparing the frames gives,
at ${\cal O} (\epsilon^0)$,
\begin{subequations}
\begin{eqnarray}
\rho(r)
&=&
\D_0(r)^2
\left(
  {\tilde\rho}({\tilde r})
+
\frac{\D_{0\Y}(r) }{2\D_0(r)}
\frac{A_t(r)^2}{f(r)}
{\tilde T}
({\tilde r})
\right),
\\
p_r(r)
&=&
\D_0(r)^2
\left(
  {\tilde p}_r ({\tilde r})
+
\frac{\D_{0\Y}(r) (r) h(r) A_r(r)^2}{2\D_0(r)}
{\tilde T}
({\tilde r})
\right),
\nonumber
\\
\\
p_t(r)
&=&
\D_0(r)^2
{\tilde p}_t
({\tilde r}),
\end{eqnarray}
\end{subequations}
and at ${\cal O} (\epsilon^1)$,
\begin{subequations} 
\begin{eqnarray}
\omega(r)&=&{\tilde \omega} ({\tilde r}),
\\
\Omega&=&{\tilde \Omega},
\\
\label{a3eq0}
a_3(r)
&=&
-
\frac{{\omega} ({r})}{f(r)}A_t(r).
\end{eqnarray}
\end{subequations}
The isotropic fluid ${\tilde p}_r({\tilde r})={\tilde p}_t ({\tilde r})$ 
in the Jordan frame
can be mapped to the anisotropic fluid  ${p}_r (r)\neq {p}_t(r)$
in the Einstein frame,
unless $\D_{0\Y}(r)=0$.
In general, 
the surface condition ${\tilde p}_r(\tilde{\R})=0$
in the Jordan frame
also corresponds to the condition $p_r (\R)=0$
in the Einstein frame with $\tilde{\R}={\tilde \R}\left({\R}\right)$
for $A_r(\R)=0$ or $\D_{0\Y}(\R)=0$.
Also, the frame-dragging function and the moment of inertia
are conformally invariant.
Rewriting the vector field $A_\mu$
in terms of the new radial coordinate ${\tilde r}$
\be
\label{el_new}
A_\mu dx^\mu
&=&
A_t({\tilde r}) dt
+A_{{\tilde r}} ({\tilde r}) d{\tilde r}
+\epsilon {\tilde r}^2{\tilde a}_3({\tilde r})\sin^2\theta d\varphi
\nonumber\\
&+&
{\cal O} 
\left(\epsilon^2\right),
\ee
where
\begin{subequations}
\be
A_{\tilde r}
({\tilde r})
&:=&\frac{A_r(r)}
              {\sqrt{\D_0(r)}\left(1+\frac{r\D_{0,r} (r)}{2\D_0(r)}\right)},
\\
{\tilde a}_3({\tilde r})
&:=&
\frac{a_3(r)}{\D_0(r)},
\ee
\end{subequations}
Eq. \eqref{a3eq0}
can be rewritten as 
\be
\label{a3eq02}
{\tilde a}_3({\tilde r})
&=&
-
\frac{{\tilde \omega} ({\tilde r})}{{\tilde f}({\tilde r})}A_t({\tilde r}).
\ee
Up to ${\cal O}(\epsilon^1)$,
the relation Eq. \eqref{a3eq0} also reads
\be
\label{induce1}
A^\varphi=\Omega A^t,
\ee
where 
$A^\varphi:=g^{\varphi \mu}A_\mu$ and $A^t:=g^{t \mu}A_\mu$,
while Eq.~\eqref{a3eq02} reads
\be
\label{induce12}
{\tilde A}^{\varphi}={\tilde \Omega} {\tilde A}^{\tilde t},
\ee
where 
${\tilde A}^\varphi:={\tilde g}^{\varphi \mu}A_\mu$
and 
${\tilde A}^t:={\tilde g}^{t \mu}A_\mu$.
Thus, the azimuthal component of
the vector field is induced 
by the rotation of the star.

The angular momenta, defined by Eq. \eqref{am},
are related by
\begin{eqnarray}
{\tilde {\cal J}}={\cal J}+\Delta {\cal J}_{V_1},
\end{eqnarray}
where
\be
\Delta {\cal J}_{V_1}
&:=&
\int 
dr d\theta d\varphi
\left(
r^2\sin\theta
\sqrt{\frac{f(r)}{h(r)}}
\right)
\frac{r^2\sin^2\theta}{f(r)}
\nonumber\\
&\times&
\left(
\frac{1}{2}
\D_0 (r)
\D_{0\Y}(r)
a_3(r)
\right)
{\tilde T}_{(m)}
({\tilde r})
A_t(r).
\ee
The moment of inertia,
defined in Eq. \eqref{moi},
are also not invariant 
\begin{eqnarray}
{\tilde {\cal I}}
={\cal I}
+\frac{\Delta {\cal J}_{V_1}}{\Omega}.
\ee
The existence of the azimuthal component 
of the vector field causes the angular momentum and
moment of inertia not to be frame-invariant.

\subsection{The case of Eq. \eqref{ar0}}
\label{sec42}

Next, we consider the case of Eq. \eqref{ar0}.
In the slow-rotation approximation, 
\begin{subequations}
\be
\E(Y)
&=&
\E_0(r)
+{\cal O} (\epsilon^2)
:=
\E
\left(\frac{A_t(t)^2}{2f(r)}\right)
+
{\cal O} (\epsilon^2),
\nonumber
\\
\\
\E_{0\Y}(r)
&:=& 
\E_{,\Y} \left(\frac{A_t(r)^2}{2f(r)}\right).
\ee
\end{subequations}
The disformal tranformation \eqref{disformalv2}
relates the radial coordinates
and the metric components
in the Jordan and Einstein frames,
respectively,
as
Eq. \eqref{tr2}
and 
\begin{subequations}
\begin{eqnarray}
{\tilde f}
({\tilde r})
&=&
\D_0(r)
\left(
f(r)-\E_0(r) A_t(r)^2
\right),
\\
{\tilde h} ({\tilde r})
&=&
h(r) \left(1+\frac{r}{2\D_0(r)}\frac{d\D_0}{dr}(r)\right)^2,
\\
{\tilde  \omega} ({\tilde r})
-
{\tilde \Omega}
&=&
 \omega(r)
-\Omega
+a_3(r) \E_0(r) A_t(r),
\end{eqnarray}
\end{subequations}
where
\begin{eqnarray}
\frac{d\D_0(r)}{dr}
&=&
-
\frac{\D_{0\Y}(r)A_t(r) \left(
-2f (r)A_t'(r)
+A_t(r)f'(r)
\right)}{2f(r)^2}.
\nonumber\\
\end{eqnarray}
In order to be compatible with the asymptotic flatness
in the Jordan frame \eqref{asym2},
we impose Eq. \eqref{asym} and 
\begin{subequations}
\label{af3}
\begin{eqnarray}
\D_0(r)
&\to& 
{\rm const}>0,
\\
A_t(r)^2
\times 
\E_0(r)
&\to& 0,
\\
r
\frac{d}{dr}
\left(A_t(r)^2\right)
\times 
\D_{0\Y}(r)
&\to & 
0,
\\
r
f'(r)
A_t(r)^2
\times 
\D_{0\Y}(r)
&\to & 
0,
\\
a_3 (r)
A_t(r)
\times \E_0(r)
&\to& 0,
\end{eqnarray}
\end{subequations}
as $r\to \infty$.
Under the conditions \eqref{af3},
the ADM masses \eqref{adms2} are shown to be invariant
\begin{eqnarray}
{\tilde M}_{\rm ADM}
=
M_{\rm ADM}.
\end{eqnarray}
Comparing frames gives, 
at ${\cal O} (\epsilon^0)$,
\begin{subequations}
\begin{eqnarray}
\rho(r)
&=&
\D_0(r)^2
\sqrt{1-\frac{\E_0(r)}{f(r)}A_t(r)^2}
\nonumber\\
&\times&
\left(
\frac{\tilde\rho}
{1-\frac{\E_0(r)}{f(r)}A_t(r)^2}
\right.
\nonumber\\
&+&
\left.
\frac{\D_0(r) A_t(r)^2}{f(r)}
\Q_0(r)
\right),
\\
p_r(r)
&=&
\D_0(r)^2
\sqrt{1-\frac{\E_0(r)}{f(r)}A_t(r)^2}
  {\tilde p}_r ({\tilde r}),
\\
p_t(r)
&=&
\D_0(r)^2
\sqrt{1-\frac{\E_0(r)}{f(r)}A_t(r)^2}
{\tilde p}_t
({\tilde r}),
\end{eqnarray}
\end{subequations}
where $\Q_0(r)$ is the ${\cal O} (\epsilon^0)$ part of 
$\Q$ defined in Eq. \eqref{calq},
\be
\Q_0(r)
&:=&
\frac{\E_{0\Y} (r)A_t(r)^2{\tilde\rho}({\tilde r})}
        {2f(r)\left(1-\frac{\E_0(r) A_t(r)^2}{f(r)}\right)}
+
\frac{\D_{0\Y}(r) {\tilde T}_{(m)} ({\tilde r})}{2\D_0(r)^2},
\nonumber\\
\ee
and at ${\cal O} (\epsilon^1)$, 
\begin{subequations} 
\begin{eqnarray}
\omega(r)&=&
\frac{{\tilde \omega} ({\tilde r})}
        {1-\frac{\E_0(r)} {f(r)} A_t(r)^2},
\\
\Omega&=&{\tilde \Omega},
\\
\label{a3eq1}
a_3(r)
&=&
-
\frac{{\omega} ({r})} {f(r)}
        A_t(r).
\end{eqnarray}
\end{subequations}
In general, 
the surface condition ${\tilde p}_r(\tilde{\R})=0$
in the Jordan frame
also corresponds to the condition $p_r (\R)=0$
in the Einstein frame with $\tilde{\R}={\tilde \R}\left({\R}\right)$.
The frame-dragging function and the moment of inertia
are conformally invariant,
and 
the last term
represents the induced magnetic component of the
vector field induced by the rotation of the spacetime.
The vector field $A_\mu$
in terms of the new radial coordinate ${\tilde r}$
can be expressed as Eq.~\eqref{el_new}
with
\begin{subequations} 
\be
A_{\tilde r}({\tilde r})&=&0, 
\\
{\tilde a}_3({\tilde r})&=&\frac{a_3(r)}{\D_0(r)},
\ee
\end{subequations}
and then Eq. \eqref{a3eq1} is identical to Eq. \eqref{a3eq0}
and can be rewritten as  Eq. \eqref{a3eq02}.
Up to ${\cal O}(\epsilon^1)$,
the relation Eq. \eqref{a3eq1} also reads
Eq. \eqref{induce1},
while
Eq. \eqref{a3eq02} reads Eq.~\eqref{induce12}.
Thus, the azimuthal component of
the vector field is induced 
from the time component of it
by the rotation of the star.

The angular momenta and moment of inertia
defined by Eqs. \eqref{am} and \eqref{moi},
respectively,
are related by
\begin{subequations}
\begin{eqnarray}
{\tilde {\cal J}}&=&{\cal J}+\Delta {\cal J}_{V_2},
\\
{\tilde {\cal I}}
&=&{\cal I}
+\frac{\Delta {\cal J}_{V_2}}{\Omega},
\end{eqnarray}
\end{subequations}
where the correction term is given by
\be
\Delta {\cal J}_{V_2}
&:=&
\int 
dr d\theta d\varphi
\left(
r^2\sin\theta
\sqrt{\frac{f(r)}{h(r)}}
\right)
\nonumber\\
&\times&
\frac{r^2
\D_0(r)^2
A_t(r)}{f(r)}
a_3(r)
\sqrt{
1-\frac{\E_0(r) A_t(r)^2}{f(r)}
}
\sin^2\theta
\nonumber\\
&\times&
\left(
\frac{\E_0 (r) {\hat\rho} (\tilde r)}
        {1-\frac{\E_0(r)}{f(r)} A_t(r)^2}
\left(
1+\frac{\E_{0\Y} (r) \D_0(r)A_t(r)^2}{2\E_0(r) f(r)}
\right)
\right.
\nonumber\\
&+&
\left.
\frac{\D_{0\Y}(r)}{2\D_{0}(r)}
{\tilde T}_{(m)} ({\tilde r})
\right).
\ee
The existence of the azimuthal component 
of the vector field causes the angular momentum and
moment of inertia not to be frame-invariant.

\section{The vector disformal transformation with the $U(1)$ gauge symmetry}
\label{sec5}

Finally, 
we consider the vector-tensor theories 
with the $U(1)$ gauge symmetry
\be
S=S_g \left[g_{\mu\nu},F_{\mu\nu}\right]
+  S_m\left[{\tilde g}_{\mu\nu},\Psi_m\right],
\ee
where the matter action is given by Eq. \eqref{matter}
and 
the relation between the two metrics is given 
by Eq.~\eqref{disformalf2}.
Up to ${\cal O} (\epsilon^1)$,
the vector field is assumed to be 
\begin{eqnarray}
A_\mu dx^\mu
=
  A_t (r)dt
+
\epsilon
r^2 a_3(r)\sin^2\theta 
d\varphi
+
{\cal O}
\left(\epsilon^2\right),
\end{eqnarray}
where
$A_t(r)$ and $a_3(r)$ are the functions of $r$.
Without loss of generality,
we may set $A_r(r)=0$
because of the $U(1)$ gauge symmetry.

We expand $\F$ and $\G$ in terms of $\epsilon$,
\begin{subequations}
\be
&&
\F
=
\F_0(r)
+{\cal O} \left(\epsilon^2\right)
:=
-\frac{2h(r)A_t'(r)^2}{f(r)}
+{\cal O} \left(\epsilon^2\right),
\nonumber
\\
\\
&&
\G
=
\G_0(r)
+{\cal O} \left(\epsilon^2\right)
:=
\frac{2h(r)^2A_t'(r)^4}{f(r)^2}
+{\cal O} \left(\epsilon^2\right),
\ee
\end{subequations}
and define 
\begin{subequations}
\be
&&
\V_{0}(r)=\V\left(\F_0(r),\G_0(r)\right),
\\
&&
\W_{0}(r)=\W\left(\F_0(r),\G_0(r)\right),
\\
&&
\V_{0\F}(r)=\V_{,\F}\left(\F_0(r),\G_0(r)\right),
\\
&&
\W_{0\F}(r)=\W_{,\F}\left(\F_0(r),\G_0(r)\right),
\\
&&
\V_{0\G}(r)=\V_{,\G}\left(\F_0(r),\G_0(r)\right),
\\
&&
\W_{0\G}(r)=\W_{,\G}\left(\F_0(r),\G_0(r)\right),
\ee
\end{subequations}
as the ${\cal O} (\epsilon^0)$ part of 
$\V$,
$\W$,
$\V_{,\F}:= \partial \V/\partial \F$,
$\V_{,\G}:= \partial \V/\partial \G$,
$\W_{,\F}:= \partial \W/\partial \F$,
and 
$\W_{,\G}:= \partial \W/\partial \G$.
Eq.~\eqref{disformalf2}
relates the radial coordinates and the metric components 
in the Jordan and Einstein frames,
respectively, by 
\be
\label{rfr}
{\tilde r}:=\sqrt{\V_0(r)}r,
\ee
and
\begin{subequations}
\begin{eqnarray}
{\tilde f}
({\tilde r})
&=&
\V_0(r)
f (r)
\left(1-\frac{h(r)\W_0(r)}{f(r)}\left(A_t'(r)\right)^2\right),
\nonumber
\\
\\
{\tilde h}
({\tilde r})
&=&
\frac{
\left(1+\frac{r}{2\V_0(r)}\frac{d \V_0}{dr}(r)\right)^2}
{1-\frac{h(r)\W_0(r)}{f(r)}\left(A_t'(r)\right)^2}
h(r),
\ee
\be
{\tilde  \omega}({\tilde r})
-
{\tilde \Omega}
&=&
 \omega (r)
-\Omega
\nonumber\\
&+&
\frac{h(r)A_t'(r)\W_0(r)}
        {r^2}
\frac{d}{dr}
\left(r^2 a_3(r)\right),
\end{eqnarray}
\end{subequations}
where
\begin{eqnarray}
\label{dvdr}
\frac{d\V_0}{dr}(r)
&=&
-
\frac{2A_t'(r)}
{f(r)^2}
\nonumber\\
&\times&
 \left[
A_t'(r)
\left(
-h (r) f'(r)
+f(r) h'(r)
\right)
\right.
\nonumber
\\
&+&
\left.
2 f(r)h(r) A_t''(r)
\right]
\nonumber\\
&\times&
\left[
\V_{0\F}(r)
-
\frac{2h(r) A_t'(r)^2}{f(r)} 
\V_{0\G}(r)
\right].
\end{eqnarray}
In order to be compatible with the asymptotic flatness
in the Jordan frame \eqref{asym2},
we impose Eq. \eqref{asym} and 
\begin{subequations}
\label{af4}
\begin{eqnarray}
\V_0(r)
&\to& {\rm const}>0,
\\
\left(A_t'(r)\right)^2
\times \W_0(r)
&\to &0,
\\
rf'(r)
\left(A_t'(r)\right)^2
\times 
\V_{0\F}(r)
&\to & 0,
\\
rh'(r)
\left(A_t'(r)\right)^2
\times 
\V_{0\F}(r)
&\to & 0,
\\
r
\frac{d}{dr}
\left[\left(A_t'(r)\right)^2\right]
\times
\V_{0\F}(r)
&\to & 0,
\\
rf'(r)
\left(A_t'(r)\right)^4
\times 
\V_{0\G}(r)
&\to & 0,
\\
rh'(r)
\left(A_t'(r)\right)^4
\times
\V_{0\G}(r)
&\to & 0,
\\
r
\frac{d}{dr}
\left[\left(A_t'(r)\right)^4\right]
\times 
\V_{0\G}(r)
&\to & 0,
\\
\frac{A_t'(r)}{r^2}
\frac{d}{dr}
\left[
 r^2  a_3(r)
\right]
\times \W_0(r)
&\to &0,
\end{eqnarray}
\end{subequations}
as $r\to \infty$.
Under the conditions \eqref{af4},
the ADM masses \eqref{adms2} are shown to be invariant,
\begin{eqnarray}
{\tilde M}_{\rm ADM}
=
M_{\rm ADM}.
\end{eqnarray}
Eq. \eqref{rel} reduces to 
\begin{eqnarray}
T_{(m)} ^{\mu\nu}
&=&
\V
\sqrt{\frac{\tilde g}{g}}
\left[
{\tilde T}_{(m)}^{\mu\nu}
-
2\Q_\F 
g^{\mu\alpha}
g^{\nu\beta}
g^{\rho\sigma}
F_{\alpha\rho} F_{\beta\sigma}
\right.
\nonumber\\
&-&
4\Q_\G
g^{\mu\kappa}
g^{\nu\omega}
g^{\alpha\beta}
g^{\rho\sigma}
g^{\gamma\delta}
 F_{\kappa\alpha} 
 F_{\rho \beta}
 F_{\sigma\gamma} 
 F_{\omega\delta}
\nonumber\\
&-&
\left.
\W
g^{\mu\kappa}
g^{\nu\omega}
F_{\kappa\alpha}
F_{\omega \beta}
{\tilde T}_{(m)}^{\alpha\beta}
\right],
\end{eqnarray}
where
\begin{subequations}
\be
&&\Q_\F
:= \frac{\V_{,\F}}{\V^2}
{\tilde T}_{(m)}
+\W_{,\F}
g^{\rho\sigma}
F_{\rho\alpha}
F_{\sigma\beta}
{\tilde T}_{(m)}^{\alpha\beta},
\\
&&
\Q_\G
:= \frac{\V_{,\G}}{\V^2}
{\tilde T}_{(m)}
+\W_{,\G}
g^{\rho\sigma}
F_{\rho\alpha}
F_{\sigma\beta}
{\tilde T}_{(m)}^{\alpha\beta}.
\ee
\end{subequations}
Comparing frames gives,
at ${\cal O} (\epsilon^0)$,
\begin{widetext}
\begin{subequations}
\begin{eqnarray}
\rho(r)
&=&
\V_0(r)
\left[
\sqrt{\frac{\tilde g}{g}}\Bigg|_0
(r)
\right]
\left\{
\frac{
{\tilde\rho}({\tilde r})
-\frac{\W_0(r) h(r) (A_t')^2}
        {f(r)}
{\tilde p}_r({\tilde r})}
{\V_0(r)\left(1-\frac{h(r)\W_0(r)}{f(r)}\left(A_t'(r)\right)^2\right)}
-
2
\left(
\frac{\Q_{0\F}(r)h(r) \left(A_t'(r)\right)^2}{f(r)}
-\frac{2\Q_{0\G}(r)h(r)^2 \left(A_t' (r)\right)^4}{f(r)^2}
\right)
\right\},
\nonumber\\
\\
p_r(r)
&=&
\V_0(r)
\left[
\sqrt{\frac{\tilde g}{g}}
\Bigg|_0
(r)
\right]
\left\{
\frac{
{\tilde p}_r({\tilde r})
-\frac{\W_0(r) h(r) \left(A_t'(r)\right)^2}
        {f(r)}
{\tilde \rho} ({\tilde r})}
  {\V_0(r)\left(1-\frac{h(r)\W_0(r)}{f(r)}(
\left(A_t'(r)\right)^2\right)}
+
2\left(
\frac{\Q_{0\F}(r)h(r) \left(A_t'(r)\right)^2}{f(r)}
-\frac{2\Q_{0\G}(r)h(r)^2 \left(A_t' (r)\right)^4}{f(r)^2}
\right)
\right\},
\nonumber\\
\\
p_t(r)
&=&
\left[
\sqrt{\frac{\tilde g}{g}}
\Bigg|_0
(r)
\right]
{\tilde p}_t
({\tilde r}),
\end{eqnarray}
\end{subequations}
\end{widetext}
where 
\begin{subequations}
\be
\sqrt{\frac{\tilde g}{g}}
\Bigg|_0
(r)
&:=&
\V_0(r)^2
\left(1-\frac{h(r)\W_0(r)}{f(r)}
\left(A_t'(r)\right)^2\right),
\\
\Q_{0\F}(r)
&:=& 
\frac{\V_{0\F} (r)}{\V_0(r)^2}
{\tilde T}_{(m)}
+
\W_{0\F}(r)
h(r)A_t'(r)^2
\nonumber\\
&\times&
\frac{
{\tilde \rho} ({\tilde r})
-{\tilde p}_r ({\tilde r})}
{\V_{0}f(r) \left(1-\frac{h(r) \W_0(r)}{f(r)} A_t'(r)^2\right)},
\nonumber
\\
\\
\Q_{0\G}(r)
&:=&
 \frac{\V_{0\G} (r)}{\V_0(r)^2}
{\tilde T}_{(m)}
+
\W_{0\G}(r)
h(r)A_t'(r)^2
\nonumber\\
&\times&
\frac{
{\tilde \rho}({\tilde r}) 
-{\tilde p}_r ({\tilde r}) }
{\V_{0}(r)f(r) \left(1-\frac{h(r) \W_0(r)}{f(r)} A_t'(r)^2\right)},
\nonumber\\
\end{eqnarray}
\end{subequations}
and at ${\cal O} (\epsilon^1)$,
\begin{subequations}
\begin{eqnarray}
&&
\omega(r)
=
\frac{{\tilde  \omega}(r)}
        {1-\frac{h(r)\W_0(r)}{f(r)}\left(A_t'(r)\right)^2},
\\
&&
\Omega=
{\tilde \Omega},
\\
\label{a3eq2}
&&
\left(r^2 a_3(r)\right)_{,r}
+\frac{r^2{\omega}({r})}{f(r)}A_t'(r)
=0.
\end{eqnarray}
\end{subequations}
In general, 
the surface condition ${\tilde p}_r(\tilde{\R})=0$
in the Jordan frame
also corresponds to the condition $p_r (\R)=0$
in the Einstein frame with $\tilde{\R}={\tilde \R}\left({\R}\right)$
only for $\W_0(\R)=\Q_{0\F}(\R)=\Q_{0\G}(\R)=0$.
Rewriting the vector field $A_\mu$
in terms of the new radial coordinate ${\tilde r}$
\be
A_\mu dx^\mu
=A_t({\tilde r}) dt
+
\epsilon {\tilde r}^2{\tilde a}_3({\tilde r})\sin^2\theta d\varphi
+{\cal O} 
\left(\epsilon^2\right),
\ee
where
\be
{\tilde a}_3({\tilde r})
&:=&
\frac{a_3(r)}{\V_0(r)},
\ee
Eq. \eqref{a3eq2}
can be rewritten as 
\be
\label{a3eq22}
\left({\tilde r}^2 {\tilde a}_3(r)\right)_{,{\tilde r}}
+\frac{{\tilde r}^2{\tilde \omega}({\tilde r})}{{\tilde f}({\tilde r})}A_{t,{\tilde r}}({\tilde r})
=0.
\ee
Up to ${\cal O}(\epsilon^1)$,
the relation Eq. \eqref{a3eq2} also reads
\be
\label{induce2}
F_r{}^{\varphi}=\Omega F_r{}^t,
\ee
where 
$F_r{}^\varphi:=g^{\varphi \mu}F_{r\mu}$
and  
$F_r{}^t:=g^{t \mu}F_{r\mu}$,
while
Eq. \eqref{a3eq22} reads
\be
\label{induce22}
{\tilde F}_{\tilde r}{}^{\varphi}={\tilde \Omega} {\tilde F}_{\tilde r}{}^t,
\ee
where 
${\tilde F}_{\tilde r}{}^\varphi:={\tilde g}^{\varphi \mu}F_{{\tilde r}\mu}$
and  
${\tilde F}_{\tilde r}{}^t:={\tilde g}^{t \mu}{F}_{{\tilde r}\mu}$.
Thus, the magnetic component of the vector field 
is induced  from the electric component of it
by the rotation of the star.

The angular momenta and moment of inertia
defined by Eqs. \eqref{am} and \eqref{moi},
respectively, are related by
\begin{subequations}
\begin{eqnarray}
{\tilde {\cal J}}
&=&{\cal J}+\Delta {\cal J}_{F},
\\
{\tilde {\cal I}}
&=&
{\cal I}
+\frac{\Delta {\cal J}_{F}}{\Omega},
\end{eqnarray}
\end{subequations}
where
\be
\Delta {\cal J}_{F}
&:=&
\int 
dr d\theta d\varphi
\left(
r^2\sin\theta
\sqrt{\frac{f(r)}{h(r)}}
\right)
\nonumber\\
&\times&
\frac{d}{dr}
\left(
r^2a_3(r)
\right)
\left(
\V_0(r)^2
A_t'(r)
\sin^2\theta
\right)
\nonumber\\
&\times&
\left[
\frac{\W_0(r)h(r)}{f(r)}
\left(
   {\tilde \rho}({\tilde r})
- {\tilde p}_r ({\tilde r})
\right)
\right.
\nonumber\\
&-&
\left.
2\V_0(r)
\left(
\Q_{0\F}(r)
\frac{h(r)}{f(r)}
-2\Q_{0\G}(r)
\frac{h(r)}{f (r)}
\left(A_t'(r)
\right)^2
\right)
\right.
\nonumber\\
&\times&
\left.
\left(
1
-\frac{h(r)\W_0(r)}
         {f(r)}
\left(A_t'(r)\right)^2
\right)
\right].
\ee
Thus, 
the difference in 
the angular momenta 
and 
moments of inertia
is given by the magnetic component of the vector field
$F_{r\varphi}$.
The existence of the magnetic component
of the vector field strength causes the angular momentum and
moment of inertia not to be frame-invariant.

\section{Conclusions}
\label{sec6}

In this paper, 
we have investigated how physical quantities associated with relativistic stars 
in the Jordan and Einstein frames are related by the generalized disformal transformations
constructed by the scalar and vector fields. 
We have considered the rigidly rotating star within the slow-rotation approximation, 
and expanded the spacetime metric and the scalar  or vector field
in terms of the small dimensionless parameter which is of the order of the dimensionless spin of the star. 
We have considered the most general scalar disformal transformation constructed by the scalar field
\eqref{disformal}, 
and by the vector field without and with the $U(1)$ gauge symmetry,
Eqs. \eqref{disformalv2} and \eqref{disformalf2},
respectively.

We have called the metric after the disformal transformation ${\tilde g}_{\mu\nu}$ 
the Jordan frame,
since in this frame
the scalar field was not directly coupled to the metric.
as in Eq. \eqref{matter}.
Correspondingly,
we called the metric before the disformal transformation ${g}_{\mu\nu}$ 
the Einstein frame.
The relation of the radial coordinates and the metric functions
could be found via the direct comparisons
of the Jordan and Einstein frames.
Similarly,
the relation of the components of 
the matter energy-momentum tensor 
could be found through Eq. \eqref{rel}.
We have required
that the metric in the Jordan frame ${\tilde g}_{\mu\nu}$
satisfies the circularity conditions \eqref{circularity2},
which in the slow-rotation limit reduced to the form of Eq.~\eqref{ansatzb}.
The circularity conditions could be automatically satisfied 
in the case of the scalar disformal transformation \eqref{disformal}
and the vector disformal transformation with the $U(1)$ symmetry \eqref{disformalf2}.
On the other hand,
in the case of the vector disformal transformation without
the $U(1)$ gauge symmetry,
Eq. \eqref{disformalv2},
the circularity conditions could be satisfied 
in the case of 
either the pure conformal coupling or the vanishing radial component of the vector field.

At the zeroth order of the slow-rotation approximation,
by imposing that the spacetime metrics $g_{\mu\nu}$ and ${\tilde g}_{\mu\nu}$
in the Einstein and Jordan frames were asymptotically flat,
we have obtained the conditions
that the metric functions and
the scalar or vector field had to be satisfied
at the spatial infinity.
With these conditions, 
we have shown
that the ADM mass is frame-invariant.

At the first order of the slow-rotation approximation, 
we have discussed the disformal transformation of the physical quantities
associated with the rotation of the star,
namely,  
the frame-dragging function, 
angular velocity, 
angular momentum, 
and
moment of inertia of the star. 
We have shown that the angular velocity of the star is frame-invariant 
in all the classes of the disformal transformation.
On the other hand, 
while the angular momentum and moment of inertia 
were invariant under the scalar disformal transformation \eqref{disformal}, 
they were not invariant under the vector disformal transformation
without and with the $U(1)$ gauge symmetry,
such as Eqs. \eqref{disformalv2} and \eqref{disformalf2},
respectively. 
Apart from the relations of the above quantities,
we have also obtained the relation between
the frame-dragging function
and the azimuthal component of the vector field
(the magnetic component of the vector field strength)
without (with) the $U(1)$ gauge symmetry,
Eq. \eqref{a3eq0} (Eq. \eqref{a3eq2}).
These relations could be interpreted
as the induction of 
the azimuthal component of the vector field
(the magnetic component of the vector field strength)
without (with) the $U(1)$ gauge symmetry,
by the slow rotation of the star,
Eq. \eqref{induce1} (Eq. \eqref{induce2}).

It would be important
to extend our analysis to the second-order of the slow rotation approximation,
which should cover the subjects of 
the tidal deformability and universal relations
\cite{Yagi:2013bca,Martinon:2014uua,Breu:2016ufb,Yagi:2016bkt,Doneva:2017jop}.
It would also be interesting to
explore our analysis to the more general class of the disformal transformation
constructed by e.g., 
higher-order derivatives, 
other field species,
and 
multiple components of the fields.
Definitively, 
these subjects would require more computations
and 
should be left for our future publications.

\acknowledgments{
M.M.~was supported by the Portuguese national fund 
through the Funda\c{c}\~{a}o para a Ci\^encia e a Tecnologia
in the scope of the framework of the Decree-Law 57/2016 
of August 29, changed by Law 57/2017 of July 19,
and the CENTRA through the Project~No.~UIDB/00099/2020.
}

\bibliography{disformal_refs}
\end{document}